\documentclass{article}
\usepackage{includes/preamble2} 

\title{Analysis of the RMM-01 Market Maker}

\author[1]{Waylon Jepsen}
\author[1]{Colin Roberts}

\affil[1]{Senior Research Engineer, Primitive Bits}
\date{\today}

\begin{document}
\maketitle

\begin{abstract}
\glspl{CFMM} are a popular market design for \glspl{DEX}. \glspl{LP} supply the \gls{CFMM} with assets to enable trades. In exchange for providing this liquidity, an \gls{LP} receives a token that replicates a payoff determined by the trading function used by the \gls{CFMM}. In this paper, we study a time-dependent \gls{CFMM} called \gls{RMM-01}. The trading function for \gls{RMM-01} is chosen such that \glspl{LP} recover the payoff of a Black--Scholes priced covered call. First, we introduce the general framework for \glspl{CFMM}. After, we analyze the pricing properties of \gls{RMM-01}. This includes the cost of price manipulation and the corresponding implications on arbitrage. Our first primary contribution is from examining the time-varying price properties of \gls{RMM-01} and determining parameter bounds when \gls{RMM-01} has a more stable price than \gls{Uniswap}. Finally, we discuss combining lending protocols with \gls{RMM-01} to achieve other option payoffs which is our other primary contribution.
\end{abstract}

\glsresetall

\section{Introduction}
\label{sec:introduction}
Blockchains are distributed computing environments that enable the construction of permissionless censorship-resistant systems such as \glspl{DEX} ~\cite{hauser_curveficurve-contract_2021, egorov_stableswap_nodate, zinsmeister_hayden, sterrett_whitepaper-rmm-01pdf_2022}. Decentralized financial instruments mitigate undesirable power distributions in existing traditional financial markets ~\cite{harvey_defi_2021}. Currently, there are billions of dollars worth of value in \gls{defi} protocols \footnote{At the time of writing, the total value locked in \gls{Uniswap} comes to \$6.77 billion.}.

\glspl{DEX} are programs that run in a decentralized computing environment that automate the function of market makers in traditional financial exchanges ~\cite{bartoletti_theory_2022}. Distributed computation has a monetary cost quantified on the Ethereum network in units of \emph{gas}. This constraint provides incentive for simpler market design. One such design is a \gls{CFMM} which is characterized by it's unique trading function ~\cite{angeris_improved_2020, angeris_constant_2021}. The trading function defines which trades are valid and assigns a price based on the current supply of reserves.

It was proven in ~\cite{angeris_replicating_2021-1} that for any concave, non-negative, non-decreasing payoff function, there exists a corresponding \gls{CFMM} replicating that payoff to \glspl{LP}. These correspondents are called \glspl{RMM}. One such example of an \gls{RMM} is the Black--Scholes covered call \gls{RMM}, currently implemented as \gls{RMM-01} by Primitive ~\cite{sterrett_primitive_2021}. This work will analyze \gls{RMM-01}.

In \cref{sec:background}, we introduce a formal framework for \glspl{CFMM} and examine the trading functions of \gls{Uniswap} and \gls{RMM-01}. Much of this background summarizes work done in ~\cite{angeris_constant_2021}. We present our main results in \Cref{sec:Properties}. In \Cref{subsec:price_impact}, we compute the price impact of a swap. Subsequently, in \Cref{sec:Manipulation} we derive the cost of price manipulation and examine the implications on arbitrage. In \Cref{sec:Compare}, we evaluate and identify parameter bounds where \gls{RMM-01} has less price impact than \gls{Uniswap}. Lastly, in \Cref{sec:Composability}, we examine a methodology for composing the \gls{RMM-01} \gls{LP} tokens with other \gls{defi} mechanisms to achieve the payoffs of other Black--Scholes priced options. 

\section{Background}
\label{sec:background}
First, we will define a \gls{CFMM} and show how the trading function is used to determine valid trades. This leads to discussion of \glspl{LP} and how they earn rewards. Since order books are a familiar concept in finance, we describe a \glspl{CFMM} that acts as an order book as a means of comparison. Options are another familiar concept which we introduce briefly before defining the \gls{CFMM} called \gls{RMM-01} that allows \glspl{LP} to achieve a payoff of a Black--Scholes \gls{CC}.

\subsection{Constant Function Market Makers}
In \gls{tradfi}, market makers facilitate the buy-and-sell orders that make up an order book and the two key actors are buyers and sellers. In \gls{defi}, trades are facilitated against the reserves of assets of a \glspl{CFMM}. For example, suppose we have a collection of $n$ assets (e.g., tokens) that can be exchanged for another. The reserves of $n$ assets $\reserves \in \R^n_+$ is called an \emph{$n$-asset liquidity pool} where for every $i\in \{1,\dots,n\}$, the quantity $R_i$ represents the quantity of asset $i$ in the pool. Reserves change when trades are executed.

Let $\deposit,\withdrawal \in \R^n_+$ be the \emph{tendered basket} and \emph{received basket}, respectively. We refer to the tuple, $(\deposit\,,\withdrawal)\in \R^n_+ \times \R^n_+$, as a \emph{proposed trade}. Specifically, $\Delta_i$ and $\Lambda_i$ denote the amount of asset $i$ tendered. Fees are typically applied to the tendered basket and they are a means of incentivizing users to provide liquidity. We define the \emph{fee} as a parameter $\gamma \in (0,1]$. 

\glsreset{CFMM}
\begin{definition}
\label{def:cfmm}
A \emph{\gls{CFMM}} is an $n$-asset pool $\R^n_+$, and a \emph{trading function} $\varphi$
\begin{equation}
    \varphi \colon \R^n_+ \to \R.
\end{equation}
Given any $\mathbf{R}$, the value $\varphi(\mathbf{R})=k$ is called the \emph{invariant}.
\end{definition}

\glspl{CFMM} have two independent actors. Actors who tender a basket of assets $\deposit$ in order to receive a nonzero basket $\withdrawal$ are \emph{swappers}. Actors who provide those assets that can be swapped are called the \emph{\glspl{LP}}. Let us first discuss swappers.
\begin{definition}
\label{def:swap}
Let $(\deposit,\withdrawal)\neq 0$, then this trade is a \emph{valid swap} if
\begin{equation}
    \label{eq:valid_trade}
    \varphi(\reserves + \gamma \deposit-\withdrawal)=\varphi(\reserves).
\end{equation}
\end{definition}
Graphically, \cref{def:swap} dictates that valid swaps move reserves along the \emph{invariant curves} where $\varphi=k$. An illustration of these curves is given in Figure \ref{fig:invariant_constant_price_curves} for both \gls{Uniswap} and \gls{RMM-01}. If $\gamma=1$, we see that a swapper is simply paying $\deposit$ in order to receive $\withdrawal$. In the case that $\gamma < 1$ (which is typical), the trade is accepted based on the discounted tendered basket, $\gamma \deposit$, but the reserves are still increased by the full $\deposit$. This remainder $(1-\gamma)\deposit$ serves to increase the value the \gls{LP}'s share of the pool which we call a \gls{LPT}. We discuss \glspl{LP} further in \Cref{subsec:liquidity_providers}. 

Suppose that we have only two assets: \textsf{Token1} and \textsf{Token2}. In \gls{tradfi}, an order book will provide the last \emph{market price} for which a trade was executed. This can be thought of as the exchange ratio $\marketprice = \frac{\Lambda_1}{\Delta_2}$. Here, we will assume that asset $i=n$ is the numeraire and that it is some stable reference such as \textsf{USDC}.
\begin{definition}
\label{def:cfmm_pricing}
Given a \gls{CFMM} the \emph{price vector} is
\begin{equation}
\pricevector \coloneqq \nabla \varphi,
\end{equation}
the \emph{reported price of asset $i$} is
\begin{equation}
    p_i \coloneqq \frac{P_i}{P_n},
\end{equation}
and the \emph{value of the reserves} is 
\begin{equation}\label{eq:v_reserves}
    V\coloneqq \frac{1}{P_n}\pricevector^T \reserves.
\end{equation}
\end{definition}

Note that since $\varphi$ is a function of the reserves, so is the reported price $p$. In ~\cite[Section 2.4]{angeris_improved_2020} the authors compute examples of pricing with two \glspl{DEX}: Balancer and Curve ~\cite{hauser_curveficurve-contract_2021, angeris_optimal_2022}. \gls{Uniswap} has the trading function $\varphi_{\textrm{Uni}}(\reserves)\coloneqq R_1 R_2$. The associated price is the ratio of the reserves $p_1 = \frac{R_2}{R_1}$. For 2-asset pools we will put $p\coloneqq p_1$. For more on Uniswap, see ~\cite{angeris_analysis_2021}. 

When a swap occurs on a \gls{CFMM}, the reserves change. If the invariant curve for a \glspl{CFMM} has nonzero curvature, the change in reserves necessarily results in a change in price. For a 2-asset pool, we can define the following.
\begin{definition}
Fix reserves $\reserves=(R_1,R_2)$ and (without loss of generality\footnote{You can achieve a swap $\Delta_2$ for $\Lambda_1$ just by change of sign when there are no fees.}) a valid swap $\Delta_1$ for $\Lambda_2$, then the \emph{price impact due to $\Delta_1$} is 
\begin{equation}
\label{eq:price_impact_def}
    g(\Delta_1) = \frac{d\Lambda_2}{d \Delta_1}.
\end{equation}
\end{definition}

The above definition appears in ~\cite{angeris_when_nodate} and more recently in ~\cite{kulkarni_towards_2022}. For example, take $\uni$ and suppose that $\Delta_1>0$, $\Delta_2=0$, and that we require $\Lambda_1 = 0$. This is reasonable since round trip trades are never profitable ~\cite[Section 2.5]{angeris_constant_2021}. Following \Cref{def:swap}, a valid swap for $\uni$ satisfies
\begin{equation}
    \uni(\reserves) = R_1 R_2 = (R_1+ \gamma \Delta_1)(R_2 - \Lambda_2) = \uni(\reserves + \gamma \deposit - \withdrawal)
\end{equation}
which implies
\begin{equation}
    \Lambda_2 = \frac{ \Delta_1 R_2}{R_1 + \Delta_1}.
\end{equation}\
Given that the price $p = \frac{R_2}{R_1}$, we have that the price impact due to $\Delta_1$ is
\begin{equation}
    g(\Delta_1)=\frac{\gamma R_1R_2}{(R_1+\gamma \Delta_1)^2}.
\end{equation}
One can define a price impact for arbitrary $n$-asset \glspl{CFMM}, but note that there is not always a unique received basket $\withdrawal$ for a given tendered basket $\deposit$.

It has been shown that depending on a price oracle can introduce centralization risk. If a price oracle is manipulated or malfunctions in any way, the financial consequences can be drastic ~\cite{tjiam_your_2021}. Note that the results of ~\cite{angeris_improved_2020} show that arbitrageurs increase market efficiency and can be utilized to mitigate oracle dependencies.


\subsection{Order Book Comparison}
\label{subsec:order_book_comparison}
In \gls{tradfi}, a common market architecture is an order book. In the order book design, buyers and sellers provide their liquidity in the form of limit orders at a specific price. Here we will examine the collection of CFMMs that describe an order book and see why they have not been used in the design of decentralized exchanges on the Ethereum network.

An order book can be built as a collection of 2-asset \glspl{CFMM} with a trading function called a \emph{constant sum trading function}. We define a generic constant sum trading function by 
\begin{equation}
\label{eq:constant_sum_market}
\varphi_{\textrm{cs}}(\mathbf{R}) = P_1R_1 + P_2R_2 = \pricevector^T \reserves,
\end{equation}
such that $P_1,P_2 \in \R_+$. The constant sum trading function dictates that valid swaps only execute at a single reported price which is akin to a limit order. The constants $P_1$ and $P_2$ define the price by
\begin{equation}
    p_{\textrm{cs}} = \frac{\partial_1 \varphi_{\textrm{cs}}}{\partial_2 \varphi_{\textrm{cs}}} = \frac{P_1}{P_2}
\end{equation}
where $\partial_i$ is the partial derivative with respect to the $i^\textrm{th}$ reserve. A visual is provided in \Cref{fig:constant_sum_market}.

When building applications on the blockchain, there is a monetary cost of computation. Thus, an on-chain order book becomes prohibitively expensive since it requires an enormous collection of constant sum markets. Instead, a \gls{CFMM} allows for a continuous range of prices using a single continuous nonlinear trading function. The dominant market architecture for \glspl{DEX} is  nonlinear \glspl{CFMM}.

\subsection{Liquidity Providers} \label{subsec:liquidity_providers}

\glspl{LP} tender assets to a pool in exchange they receive a quantity of \glspl{LPT} representing their share of pool ownership. The \gls{LPT} can always be exchanged for reserve assets if an \gls{LP} wishes to exit their position. Upon exiting, an \gls{LP} receives a basket $\withdrawal$ for their \gls{LPT}. Their share is defined by their initial tendered basket $\deposit$ and the accumulated swap fees. For more detail, see \cite{angeris_constant_2021}. We refer to the actions an \gls{LP} performs as \emph{liquidity changes}.

Swappers and \glspl{LP} have a symbiotic relationship so long as $\gamma \neq 1$. For any swap $(\deposit,\withdrawal)$ there will be $(1-\gamma)\deposit$ placed into the pool. Hence, swaps increase the amount of the underlying tokens that an \gls{LPT} is worth. More detail about the change in value for an \gls{LPT} is given in ~\cite[Section 2.6]{angeris_constant_2021}.

In \glspl{CFMM}, the reported price is updated through arbitrage. To capitalize on an arbitrage opportunity, the arbitrageurs must execute swaps on two exchanges and each swap executed by the arbitrageur on a \gls{CFMM} generates fee paid to \glspl{LP}. This mechanism drives \gls{RMM-01} to replicate the payoff of a \gls{CC}. We do not go into the exact detail here, but instead we refer the reader to the original papers ~\cite{angeris_replicating_2021, sterrett_whitepaper-rmm-01pdf_2022} that describes this trading function and its implementation. 

A \gls{LP} executes a trade of the form $(\deposit,0)$ or $(0,\withdrawal)$ that modifies the invariant $k$ but not the reported price\footnote{\glspl{LP} and swappers are orthogonal actors in a \gls{CFMM}. The fee $\gamma$ can be considered as a ``rotation'' that converts the impact of a swapper to an \gls{LP} geometrically.}. Specifically, \glspl{LP} must provide or remove liquidity along directions that preserve the gradient of the trading function. 
\begin{definition}
A \emph{valid liquidity change} is a trade $(\deposit\,,0)$ or $(0,\withdrawal)$ such that
\begin{equation}
p(\reserves) = p(\reserves+\deposit) = p(\reserves-\withdrawal)
\end{equation}
\end{definition}
The liquidity changes occur along the \emph{constant price curves} that can be seen in Figure \ref{fig:invariant_constant_price_curves} for both \gls{Uniswap} and \gls{RMM-01}.

\subsection{Black-Scholes Option Pricing}\label{Black--Scholes}
The price of a Black--Scholes option depends on the chosen \emph{strike price}, $K$, and the time until expiration $\tau$~\cite{black_pricing_2019}. Once these parameters are chosen, the option's price changes due to changes in market price $S$ of the underlying asset, the \emph{implied volatility} $\sigma$ of the underlying~\footnote{While implied volatility is variable in traditional options, it is constant for \gls{RMM-01} which results in effectively no vega.}, and the \emph{time to expiry} $\tau$ of the contract\footnote{Black--Scholes assumes a risk-free interest rate ~\cite{black_pricing_2019}. We omit this feature.}. 

The value of an option can be given in terms of the variables $d_1$ and $d_2$ defined as
\begin{equation}
   d_1(\marketprice,\tau)= \frac{\ln(\frac{\marketprice}{K}) + \frac{1}{2}\sigma^2 \tau}{\sigma\sqrt{\tau}} \qquad \textrm{and} \qquad d_2(\marketprice,\tau) = d_1 - \sigma\sqrt{\tau}.
\end{equation}
Note that throughout this paper, we denote the zero-mean unit-variance Gaussian probability density function (PDF), cumulative distribution function (CDF), and quantile functions by $\phi$, $\Phi$, and $\Phi^{-1}$, respectively.
\begin{definition}
\label{def:call_covered_call}
The \emph{value of a long call} is
\begin{equation}
    \label{eq:value_of_call}
    \vcall(\marketprice, \tau) = \marketprice \Phi(d_{1}) - K\Phi(d_{2})
\end{equation}
and the \emph{value of a covered call} is
\begin{equation}
    \label{eq:value_of_covered_call}
    \vcc(\marketprice, \tau) = \marketprice \Phi(-d_1) + K\Phi(d_2).
\end{equation}
\end{definition}

Black--Scholes priced options satisfy a put-call parity ~\cite{klemkosky_put-call_1979} that allows for a put to be priced immediately. The \emph{value of a long put} is
\begin{equation}
\label{eq:value_of_put}
    \vput(\marketprice, \tau) = \vcall - \marketprice + K = K\Phi(-d_2)-\marketprice \Phi(-d_1).
\end{equation}

\subsection{RMM-01}
\label{subsec:RMM-01}
In the paper \emph{Replicating Market Makers}~\cite{angeris_replicating_2021}, the authors provide an example of a \gls{CFMM} that approximates the payoff of a Black--Scholes \gls{CC} option. This trading function was deployed on the Ethereum main-net under the name \gls{RMM-01}~\cite{sterrett_whitepaper-rmm-01pdf_2022}. At the time of writing, the average total value locked on \gls{RMM-01} is \$550,965.38 and the average daily swap volume is \$11,058.58 ~\cite{evan_primitivefinancepool-analytics_2022}.

The behavior of \gls{RMM-01} is unlike options in \gls{tradfi}. Specifically, it is not a means to buy or sell options contracts as it does not act as a market counterparty. An \gls{RMM-01} pool is initialized when the \gls{LP} chooses a strike price $K$ (of \textsf{ETH} in \textsf{USDC}), an implied volatility $\sigma$, the time to expiry $\tau$, and adds liquidity to the pool.

Once a pool is created, other \glspl{LP} are free to provide liquidity to existing pools. Swappers (commonly arbitrageurs) enticed by profit opportunities from pricing discrepancies between \gls{RMM-01} and another market. Since \glspl{LP} grow their position through swap fees, arbitrage is a necessary ingredient for accurate replication of the Black--Scholes \gls{CC} payoff that \gls{RMM-01} claims. As an added benefit, this methodology eliminates the dependence upon an options market counterparty.

Consider the 2-asset pool with the \gls{RMM-01} trading function given by 
\begin{equation}\label{eq:tradingfunc}
    \rmm \coloneqq -K\Phi(\Phi^{-1}(1-R_1)-\sigma\sqrt{\tau}) + R_2.
\end{equation}
\textsf{Token1} is typically called the \emph{risky asset} (or the \emph{underlying} in \gls{tradfi} literature). \textsf{Token2} is called the \emph{stable asset} and is our numéraire. Note that $\rmm$ is evaluated on a per-\gls{LPT} basis and the invariant $k$ is not a measure of total liquidity as in \gls{Uniswap}. Instead, $k$ measures how accurate the replication of the payoff of a \gls{CC} is being achieved. We say that there is \emph{perfect replication} if $k=0$. The pool is over replicating if $k>0$ and under-replicating if $k<0$. We can see how the invariant curves for perfect replication change over time in Figure \ref{fig:time_dependence}.

\section{Properties of RMM-01}
\label{sec:Properties}
Here, we apply some of the definitions from \Cref{sec:background} to perform analysis on the \gls{RMM-01} trading function. We first compute the price impact and find the cost of manipulation and associated arbitrage implications. Furthermore, we compare our results to \gls{Uniswap} and introduce a bound on $\sigma\sqrt{\tau}$ in which the trading is preferable on \gls{RMM-01}. Lastly, we introduce a mechanism to construct additional Black--Scholes priced payoffs.

\Cref{eq:tradingfunc} implies that $R_1 \in (0,1)$ and if we assume perfect replication it must be that $R_2 \in (0,K)$ when $\tau>0$. Applying \Cref{def:cfmm_pricing} to the trading function in \Cref{eq:tradingfunc}, the reported price of \textsf{Token1} in terms of \textsf{Token2} for an \gls{RMM-01} pool
\begin{align}
    \label{eq:rmm-01_price}
    p = Ke^{\Phi^{-1}(1 -R_1)\sigma\sqrt{\tau}-\frac{1}{2}\sigma^{2}\tau}.
\end{align}
An \gls{LP} deposits a collection of \textsf{Token1} and \textsf{Token2} in proportional quantities so that the reporting price $p$ from \Cref{eq:rmm-01_price} matches the market price $\marketprice$. For instance, suppose a user has 1 \textsf{ETH} (\textsf{Token1}) and that the market price for \textsf{ETH} in \textsf{USDC} (\textsf{Token2}) is $S=1000$. Then the reserves of \textsf{ETH} per \gls{LPT} are given by
\begin{equation}
    \label{eq:R1_from_price}
    R_1 = 1 - \Phi \left( \frac{\ln \left(\frac{S}{K}\right) + \frac{1}{2} \sigma^2 \tau}{\sigma \sqrt{\tau}} \right),
\end{equation}
which implies that the user will deposit $R_1$ \textsf{ETH}. The user can always purchase more or fractional amounts of \glspl{LPT}. Given the user's chosen $K$, $\sigma$, and $\tau$, the number of \textsf{USDC} tokens to be deposited is
\begin{equation}
    \label{eq:R2_from_price}
    R_2 = K \Phi \left( \frac{\ln \left(\frac{S}{K}\right) - \frac{1}{2} \sigma^2 \tau}{\sigma \sqrt{\tau}} \right).
\end{equation}
As long as $\sigma \sqrt{\tau} >0$, it follows that $KR_1+R_2 <K$. This means the user may have remaining funds and can choose to purchase more \glspl{LPT}.

Applying \cref{eq:v_reserves} to \cref{eq:tradingfunc} and assuming $p=S$, we can see the value of \glspl{RMM-01} \gls{LPT} $\vrmm$ in terms of the numeraire $(n=2)$ matches that value of a \gls{CC}
\begin{equation}\label{eq:RMM-01 price vector}
\begin{aligned}
    \vrmm &= S \Phi(- d_1 ) + K \Phi ( d_2)=\vcc.
\end{aligned}
\end{equation}

Briefly, note that when $\tau = 0$ (i.e., the pool expired), the trading function becomes a constant sum market
\begin{equation}
   KR_1+R_2=K
\end{equation}
and the reported price is
\begin{align}
    p \vert_{\tau =0} = K.
\end{align}
Recall that a constant sum market is equivalent to a limit order and the geometry is seen in \Cref{fig:constant_sum_market}.

\subsection{Price Impact}
\label{subsec:price_impact}

Since \gls{RMM-01}'s trading function changes over time, it follows that the price impact does as well. Also, as \gls{RMM-01} is not symmetric, we must consider price impacts of both \textsf{Token1} and \textsf{Token2} separately. We write $g(\Delta_1\,,0)$ and $g(0\,,\Delta_2)$ to denote the price impacts due to tendering $\Delta_1$ and $\Delta_2$, respectively.

First, if the swapper tenders $\Delta_1$ then \Cref{eq:rmm-01_price} implies that the new price $g(\Delta_1,0)$ is
\begin{equation}
    \label{eq:new_price_from_delta_1}
    g(\Delta_1,0) = Ke^{\Phi^{-1}(1 -R_1-\gamma \Delta_1)\sigma\sqrt{\tau}-\frac{1}{2}\sigma^{2}\tau}.
\end{equation}
At expiry (when $\tau=0$), we have
\begin{equation}
    \label{eq:expiry_lambda_2}
    \Lambda_2 = \frac{\gamma \Delta_1}{K},
\end{equation}
which implies that swappers can only exchange assets at the strike price when the pool expires. If $\marketprice>K$, then all traders are incentivized to convert the pool into \textsf{Token2}, so an \gls{LP} recovers their \gls{LPT} payoff in \textsf{Token2}. The opposite is also valid, so if $\marketprice < K$, the pool expires with only \textsf{Token1}. 

If instead, a swapper tenders $\Delta_2$, then we can find that the new price is
\begin{equation}
    \label{eq:new_price_from_delta_2}
    g(0,\Delta_2) = Ke^{\Phi^{-1} \left(\frac{R_2 + \gamma \Delta_2}{K} \right) + \frac{1}{2} \sigma^2 \tau}.
\end{equation}
Once again, at expiry $p'\vert_{\tau=0}=p\vert_{\tau=0}$ and we find 
\begin{equation}
    \Lambda_1 = \gamma K \Delta_2.
\end{equation}

\subsection{Manipulation and Arbitrage}
\label{sec:Manipulation}

Suppose that we want to manipulate the reported price from $p\mapsto (1+\epsilon)p$, then we need to find a corresponding $\Delta_1$ or $\Delta_2$ that would lead to this price change. If the swapper wants to tender $\Delta_1$ we use \Cref{eq:rmm-01_price,eq:new_price_from_delta_1} to see
\begin{equation}
\Delta_1 = \gamma^{-1} \left( 1- R_1 - \Phi\left( \Phi^{-1}(1-R_1)+\frac{\ln(1+\epsilon)}{\sigma \sqrt{\tau}} \right) \right).
\end{equation}

If $\Delta_2$ is tendered then we use \Cref{eq:rmm-01_price,eq:new_price_from_delta_2} to get
\begin{equation}
    \Delta_2 = \gamma^{-1} K \Phi\left( \Phi^{-1}(1-R_1)-\sigma \sqrt{\tau} + \frac{\ln(1+\epsilon)}{\sigma \sqrt{\tau}} \right)-\gamma^{-1} R_2.
\end{equation}

Note that $\Delta_1$ and $\Delta_2$ are per \gls{LPT}, thus this cost increases linearly with liquidity. Arbitrageurs also need to execute a swap in order to have the reported price $p$ match an external market price $\marketprice$ to maximize their profits. In order for an arbitrageur to initiate a profitable swap, the reported price $p$ must satisfy the following bounds that depend on the fee:
\begin{equation}
    \label{eq:arbitrage_bounds}
   \gamma \marketprice \leq  p \leq \gamma^{-1} \marketprice.
\end{equation}
These bounds are true for all \glspl{CFMM}. Given $p$ satisfies the bounds in \Cref{eq:arbitrage_bounds}, they can supply $\Delta_1$ or $\Delta_2$ from \Cref{eq:new_price_from_delta_1,eq:new_price_from_delta_2} by letting $\epsilon = \gamma^{-1}-1$ if the reported price needs to increase and $\epsilon=\gamma-1$ if the price needs to decrease.

If the invariant curves for a 2-asset trading function have small curvature, then the price impact is also small. More detail on the relationship of curvature to capital efficiency is shown in ~\cite{chitra_note_2021, angeris_improved_2020}. 

\subsection{Compare with Uniswap V2}
\label{sec:Compare}
By comparing \gls{RMM-01} with \gls{Uniswap}, we can identify explicit bounds for $\sigma\sqrt{\tau}$ such that the price impact for a swap on \gls{RMM-01} is less than on \gls{Uniswap}. This proves to be useful for users wanting to trade large quantities of assets who don't want to be exposed to a large price tolerance commonly known as \emph{slippage}. Existence of these bounds is due to the fact that the curvature of the invariant curves approaches zero for \gls{RMM-01} as the pool approaches expiry (see Figure \cref{fig:time_dependence}). 
We studied price impact in \Cref{sec:Manipulation} and in this subsection we examine the price impact with respect to an infinitesimal swap. \gls{RMM-01} and \gls{Uniswap} have different trading function and consequently price assets differently. Furthermore, \gls{RMM-01} considers reserves on a per \gls{LPT} basis. Suppose that \gls{RMM-01} and \gls{Uniswap} both report price $p$ for \textsf{Token1}. How do the infinitesimal changes in price for each \gls{CFMM} relate to one another? In particular, for what choices of $\sigma$ and $\tau$ can \gls{RMM-01} achieve less price impact? 

Recall that the reported price for \gls{Uniswap} is $p_{\textrm{uni}}=\frac{R_2}{R_1}$. For any swap tendering either asset, the reserves must be a point along the same invariant curve (see \Cref{fig:invariant_constant_price_curves}). Hence, we can compute a directional derivative of the price along invariant curves to get an infinitesimal price impact. Assuming normalized reserves, $\pricevector_{\textrm{Uni}} = \begin{pmatrix} R_2 & R_1 \end{pmatrix}^T$ which is orthogonal to the invariant curves. We can apply a rotation by $\pi/2$ via a linear transformation $J$ to $\pricevector_{\textrm{Uni}}$ yields a vector tangent to the invariant curves $J\pricevector_{\textrm{Uni}} = \begin{pmatrix} -R_1 & R_2 \end{pmatrix}^T$. Computing the directional derivative along $J\pricevector_{\textrm{Uni}}$ yields
\begin{equation}
    \label{eq:uni_infinitesimal_price_impact}
    \nabla_{J\pricevector_{\textrm{Uni}}} p_{\textrm{uni}} = -R_1 \frac{\partial p_{\textrm{Uni}}}{\partial R_1} + R_2 \frac{\partial p_{\textrm{Uni}}}{\partial R_2} = 2p_{\textrm{Uni}}.
\end{equation}
By the same argument for \gls{RMM-01}, we get 
\begin{equation}
    \label{eq:rmm-01_infinitesimal_price_impact}
   \nabla_{J\pricevector_{\textrm{RMM-01}}} p_{\textrm{RMM-01}} = \frac{p_{\textrm{RMM-01}} \sigma \sqrt{\tau}}{\phi(\Phi^{-1}(1-R_1))}.
\end{equation}

If we assume $p_{\textrm{Uni}}=p_{\textrm{RMM-01}}$, then we can compare \Cref{eq:uni_infinitesimal_price_impact,eq:rmm-01_infinitesimal_price_impact} and determine that for the infinitesimal price impact for \gls{RMM-01} to be less than that of Uniswap at the same price, it must be that
\begin{equation}
\label{eq:price_impact_bounds}
    \sigma \sqrt{\tau} < 2\phi(\Phi^{-1}(1-R_1)),
\end{equation}
where $R_1$ is the reserves for \gls{RMM-01} given by \Cref{eq:R1_from_price} and $\phi$ is the standard normal PDF. We can visualize \cref{eq:price_impact_bounds} in \Cref{fig:price_impact_bounds}.

\subsection{Composability}
\label{sec:Composability}

A natural question is: given the \gls{RMM-01} \gls{LPT} can replicate the payoff of a \gls{CC}, can one obtain other Black--Scholes priced options from this \gls{LPT}? We examine the composition of assets and mechanisms that result in the payoff of a long call $\vcall(S,\tau)$ and consequently a long put. This work expands on some of the results in ~\cite{sterrett_replicating_2022} which introduces additional composability results from \gls{RMM-01}.

We propose a basic borrowing and lending mechanism that can achieve the payoff of a long call or put using \gls{RMM-01}. First, we outline the approach and then we analyze the inherent risk and maximal downsides. Our results illustrate the differences between a traditional call and one built by composing borrowing with \gls{RMM-01} \glspl{LPT}. This is a type of trade that can be implemented on blockchains today with no need for a new lending protocol.

In \gls{tradfi}, put and call options are commonly used to mitigate risk ~\cite{froot_risk_1993} and they can play a similar role in \gls{defi}. As of March 2022, the derivative market on centralized exchanges for cypto assets represents 62.8\% of trading volume~\cite{webb_cryptocompare_exchange_review_2022_03_vf-2pdf_2022} expressing an apparent demand.

Assuming perfect replication of \gls{RMM-01}, the value of an \gls{RMM-01} \gls{LPT} is given by $\vcc$ found in \Cref{eq:value_of_covered_call}. Since a \gls{CC} is equivalent to a long position in the underlying and short a call, we get the relationships between $\vcc$ and $\vcall$ seen in \Cref{def:call_covered_call}. Using the same relationship, we can achieve the payoff of a call $\vcall$ by shorting a \gls{CC} and adding a long position on the underlying. In terms of \gls{RMM-01}, we must have a method to short the \gls{LPT} and hold one unit of \textsf{Token1}.

To short an asset in \gls{defi}, one can borrow the asset with collateral and sell it. This sale can be done on a \gls{DEX} for either \textsf{Token1} or \textsf{Token2}\footnote{Could be for any other token to create a more exotic position.}. Note that the borrowed asset can be bought back at a later time or the borrowing position can be liquidated. Decentralized lending protocols also ensure at least a one-to-one loan-to-value ratio for collateral and dipping below this ratio causes liquidation. Each of these steps will be called a \emph{transaction} (labeled $\mathbf{tx}$) and are executed atomically. On a blockchain, collections of transactions are listed on\emph{blocks}.

For the remainder of this section, assume that $p=\marketprice$. Let $\tstart$ be the time of entering the position and $\tend$ be time of exiting the position such that $\tstart>\tend\geq0$. Suppose that a user provides \textsf{Token1} (e.g., \textsf{ETH}) as collateral in order to borrow an \gls{LPT}. The user can then immediately sell the \gls{LPT} for \textsf{Token1}. Because of atomic execution of transactions, we can assume both transactions occur at $\tstart$. At $\tend$, the user can buy back the \gls{LPT} and repay their debt for their original collateral and exit the position unless they have already experienced liquidation. We refer to this position as a \emph{\gls{RMM-01} synthetic call} and denote the payoff by $\synthcall$. Thus the steps to construct the $\synthcall$ from the \gls{RMM-01} \gls{LPT} are summarized as follows:

\begin{figure}[h!]
    \centering
    \begin{tikzpicture}[thick]
    \node [draw,
        rounded corners=0.2cm,
        text width=3.1cm,
        align=center,
        minimum width=3.1cm,
        minimum height=2.6cm,
    ] (A) {\vspace*{-0.2cm}\textbf{Block:} $\tstart$
    \line(1, 0){\textwidth}
    \begin{flushleft}
    \footnotesize$\mathbf{tx}_1$: Provide collateral with \textsf{Token1}
    \\\footnotesize$\mathbf{tx}_2$: Borrow one \gls{LPT} and sell one \gls{LPT} for \textsf{Token1}
    \end{flushleft}};
    \node[right=0.5cm of A,
        align=left](C) {$\dots$};
    \node [draw,
        rounded corners=0.2cm,
        text width=3.1cm,
        align=center,
        minimum width=3.1cm,
        minimum height=2.6cm,
        right=0.5cm of C
    ] (B){\vspace*{-0.2cm}\textbf{Block:} $\tend$
    \\\line(1, 0){\textwidth}
    \\\begin{flushleft}
    \footnotesize$\mathbf{tx}_1$: Purchase an \gls{LPT} using \textsf{Token1}
    \\\footnotesize$\mathbf{tx}_2$: Repay the loan to recover collateral \textsf{Token1}
    \end{flushleft}
};
    \draw[->] (A) -- (C);
    \draw[->] (C) -- (B);
    \end{tikzpicture}
    \caption{Entering the position of a $\synthcall$ at \textbf{Block:} $\tstart$ and exiting it at block \textbf{Block:} $\tend$.}
    \label{fig:rmmc}
\end{figure}
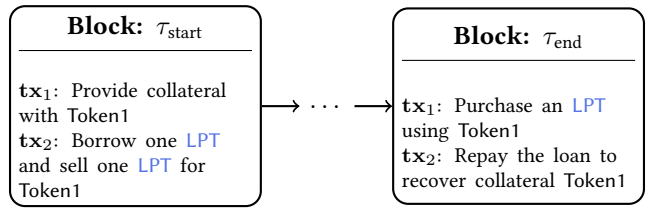

Note that the position $\synthcall$ described above is built under the assumptions of a one-to-one loan to value ratio. In \Cref{tab:collateral_payoff} below, the first column denotes $x$ as the sale price of $\vrmm$ in terms of \textsf{Token1} at $\tstart$, the second column is the chosen collateral $y\geq x$ provided to borrow the \gls{LPT}, the third column is the total amount of \textsf{Token1} the user is exposed to, the fourth column is the value of $\synthcall$ given a market price $S$, and the final column represents the differences between a $\synthcall$ and a $\vcall$.

\begin{table}[ht]
\centering
\caption{The positions depending on the \gls{CC} parameters.}
\renewcommand{\arraystretch}{1.5}
\begin{tabular}{ | c | c | c | c | c |}
\hline
Sale & Col. & \textsf{Token1} &  $\synthcall$  & $\vcall - \synthcall$\\ 
\hline
$x$ & $y$ & $x+y$ & $(x+y)S-\vrmm$ & $(x+y-1)S$\\
\hline
\end{tabular}
  \label{tab:collateral_payoff}
\end{table}
 
At $\tstart$ it is very possible that $x+y-1\neq 0$. One such case is if the user over-collateralize ($y>x$). However, the user can still achieve the long call payoff by adjusting their exposure to \textsf{Token1}. Specifically, the user needs to find an amount $z$ of \textsf{Token1} such that $x+y+z-1=0$. The key assumptions are that there is sufficient liquidity of \glspl{LPT} on a lending market and a \gls{DEX}. The cost of entering the position is the collateral $y$ equal in value to $\vrmm$ at $\tstart$. 

To achieve the payoff of a put, the user replaces all instances of \textsf{Token1} in Figure \cref{fig:rmmc} with \textsf{Token2}. This is a consequence of put-call parity. Using \cref{eq:value_of_put} we can see that $\vput = -\vcc + K$. In this case, the user may not achieve the correct $+K$ term need due to the value of \gls{LPT} at $\tstart$. 

 A loan must have at least a loan-to-value ratio of one which is why it must be that $y\geq x$. The user may provide $y>x$ to ensure that their loan-to-value ratio does not dip below one even with adverse price action of \textsf{Token1}. In essence, the difference $y-x$ can be thought of as a choice of stop-loss. By providing $y=1$, the user can ensure they will never face liquidation since the \gls{LPT} can trade for at most one of \textsf{Token1}.

Suppose that the price of \textsf{Token1} decreases to the point where the \gls{LPT} expires containing one of \textsf{Token1}. This is the worst case scenario for the user long $\synthcall$. Then, along the way, the user may find that their loan to value ratio drops below one and they will be liquidated. That is, they will necessarily forfeit their $y$ \textsf{Token1}. They will, however, still keep their $x$ \textsf{Token1} and can decide what to do from there. At any rate, the maximal loss the user experiences is $yS_{\tstart}$ where $S_{\tstart}$ is the initial price of \textsf{Token1} in terms of \textsf{Token2}. Once again, this implies the worst possible risk is $S_{\tstart}$ if the user collateralizes using $y=1$.

When an actor borrows assets from a \gls{defi} protocol, they pay an interest rate on their loan for the time they are borrowing the assets. This means that there is an additional payoff to the lender and an additional cost to the borrower that we do not consider here. 

\section{Conclusion}
In this work, we studied the time-dependent \gls{CFMM} called \gls{RMM-01} which replicates the payoff of a Black--Scholes \gls{CC}. Specifically, we computed the price impact due to swaps, discussed price manipulation and arbitrage, and we compared these to known bounds for another \gls{CFMM} called \gls{Uniswap}. Finally, we describe how to achieve the payoff of a long call using an external lending protocol. We do, however, have further open questions to answer.

\paragraph{Optimal Fees for \gls{RMM-01}.} It has been shown that fixed fees in the \gls{RMM-01} protocol do not yield perfect replication of a \gls{CC} payoff~\cite{noauthor_graphiql_nodate}. Since \gls{RMM-01} has decreasing price impact over time, we believe that the optimal fee mechanism should also decrease over time to allow for larger swaps. Preliminary work has been done here~\cite{milionis_automated_2022}.

\paragraph{Symmetric Asset Pools.} Currently \gls{RMM-01} supports \textsf{ETH} and \textsf{USDC} but it can interact with all \gls{ERC20} tokens. One potentially interesting research question is to examine the behavior of \gls{RMM-01} when the pool consists of two stable tokens. The trading function used by Curve ~\cite{noauthor_curve_2022} is often used for stable-stable swapping since there is less price impact. Does \gls{RMM-01}'s time-dependent replication of a \gls{CC} add any benefit for a stable-stable pool? Similar to the previous thought, what are the economic implications of a pool consisting of two volatile assets such as \textsf{ETH} and \textsf{BTC} pool?


\bibliographystyle{unsrt}
\bibliography{includes/main.bib}


\appendix

\section{CFMM Geometry}

\begin{figure}[h]
\centering
\def\svgwidth{\columnwidth}
    \resizebox{\columnwidth}{!}{
\begingroup%
  \makeatletter%
  \providecommand\color[2][]{%
    \errmessage{(Inkscape) Color is used for the text in Inkscape, but the package 'color.sty' is not loaded}%
    \renewcommand\color[2][]{}%
  }%
  \providecommand\transparent[1]{%
    \errmessage{(Inkscape) Transparency is used (non-zero) for the text in Inkscape, but the package 'transparent.sty' is not loaded}%
    \renewcommand\transparent[1]{}%
  }%
  \providecommand\rotatebox[2]{#2}%
  \newcommand*\fsize{\dimexpr\f@size pt\relax}%
  \newcommand*\lineheight[1]{\fontsize{\fsize}{#1\fsize}\selectfont}%
  \ifx\svgwidth\undefined%
    \setlength{\unitlength}{338.60581285bp}%
    \ifx\svgscale\undefined%
      \relax%
    \else%
      \setlength{\unitlength}{\unitlength * \real{\svgscale}}%
    \fi%
  \else%
    \setlength{\unitlength}{\svgwidth}%
  \fi%
  \global\let\svgwidth\undefined%
  \global\let\svgscale\undefined%
  \makeatother%
  \begin{picture}(1,0.63213817)%
    \lineheight{1}%
    \setlength\tabcolsep{0pt}%
    \put(0,0){\includegraphics[width=\unitlength,page=1]{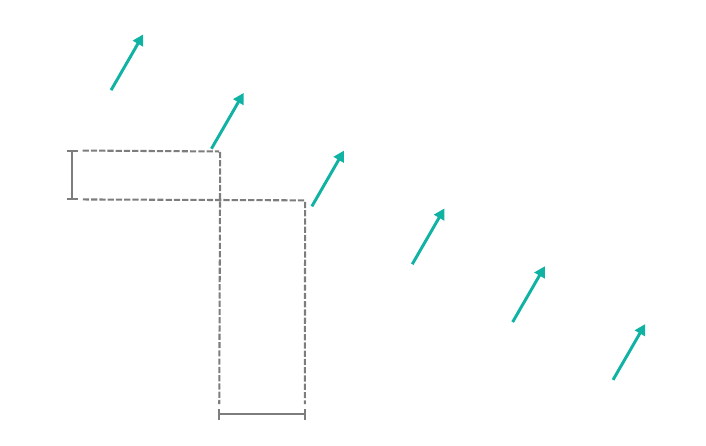}}%
    \put(0.03968649,0.37771712){\color[rgb]{0.49019608,0.49019608,0.49019608}\makebox(0,0)[lt]{\lineheight{1.25}\smash{\begin{tabular}[t]{l}$P_1$\end{tabular}}}}%
    \put(0.35676906,0.00555826){\color[rgb]{0.49019608,0.49019608,0.49019608}\makebox(0,0)[lt]{\lineheight{1.25}\smash{\begin{tabular}[t]{l}$P_2$\end{tabular}}}}%
    \put(0.53248968,0.42878687){\color[rgb]{0.0627451,0.70196078,0.63921569}\makebox(0,0)[lt]{\lineheight{1.25}\smash{\begin{tabular}[t]{l}$\pricevector_{\textrm{cs}} = \frac{\nabla \varphi}{|\nabla \varphi|}$\end{tabular}}}}%
    \put(0,0){\includegraphics[width=\unitlength,page=2]{constant_sum_market.pdf}}%
    \put(0.54408765,0.02693275){\color[rgb]{0.21176471,0.61960784,1}\makebox(0,0)[lt]{\lineheight{1.25}\smash{\begin{tabular}[t]{l}$\varphi_{\textrm{cs}} = k$\end{tabular}}}}%
    \put(0.71750272,0.02686013){\color[rgb]{0.21176471,0.61960784,1}\makebox(0,0)[lt]{\lineheight{1.25}\smash{\begin{tabular}[t]{l}$\varphi_{\textrm{cs}} = k'$\end{tabular}}}}%
  \end{picture}%
\endgroup%
}
\caption{Limit Order as a Constant Sum Market.}
\label{fig:constant_sum_market}
\end{figure}

Let us describe the geometry of \gls{CFMM} trading curves briefly. First, we can see a visual for the invariant curves for different values of $k$ in the constant sum market trading function $\varphi_{\textrm{cs}}$ described by \Cref{eq:constant_sum_market}. Alongside this, the price vector $\pricevector_{\textrm{cs}}$ is visualized as a vector field. Since the price vector is just a normalized gradient of the trading function $\varphi_{\textrm{cs}}$, $\pricevector_{\textrm{cs}}$, it is necessarily orthogonal to the invariant curves. All of this can be seen in \Cref{fig:constant_sum_market}.

\begin{figure*}[!ht]
     \centering
     \begin{subfigure}[b]{0.45\textwidth}
         \centering
         \includegraphics[width=\textwidth]{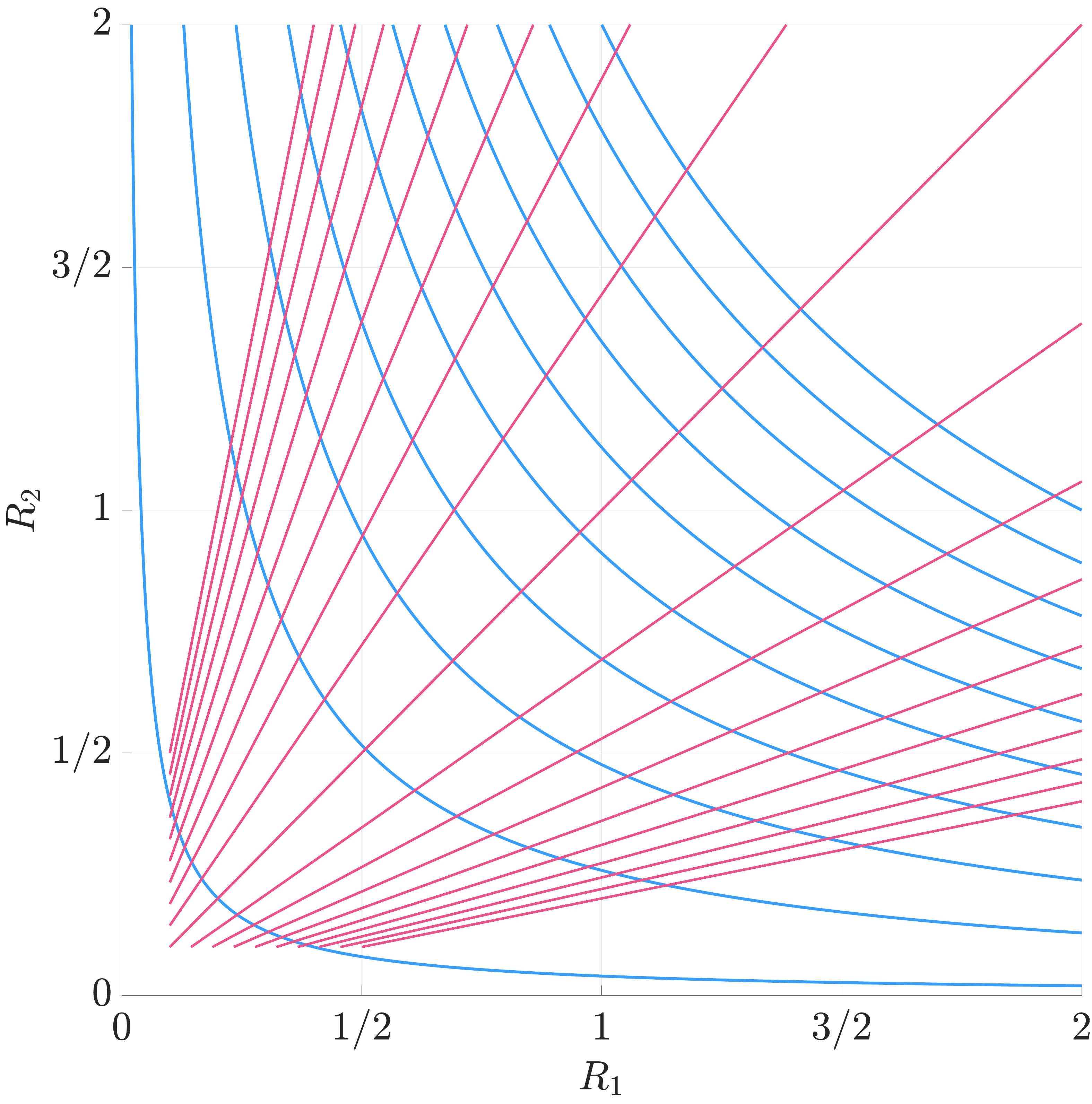}
         \caption{Curves for Uniswap V2.}
         \label{fig:uniswap_trading_function}
     \end{subfigure}
     \hfill
     \begin{subfigure}[b]{0.45\textwidth}
         \centering
         \includegraphics[width=\textwidth]{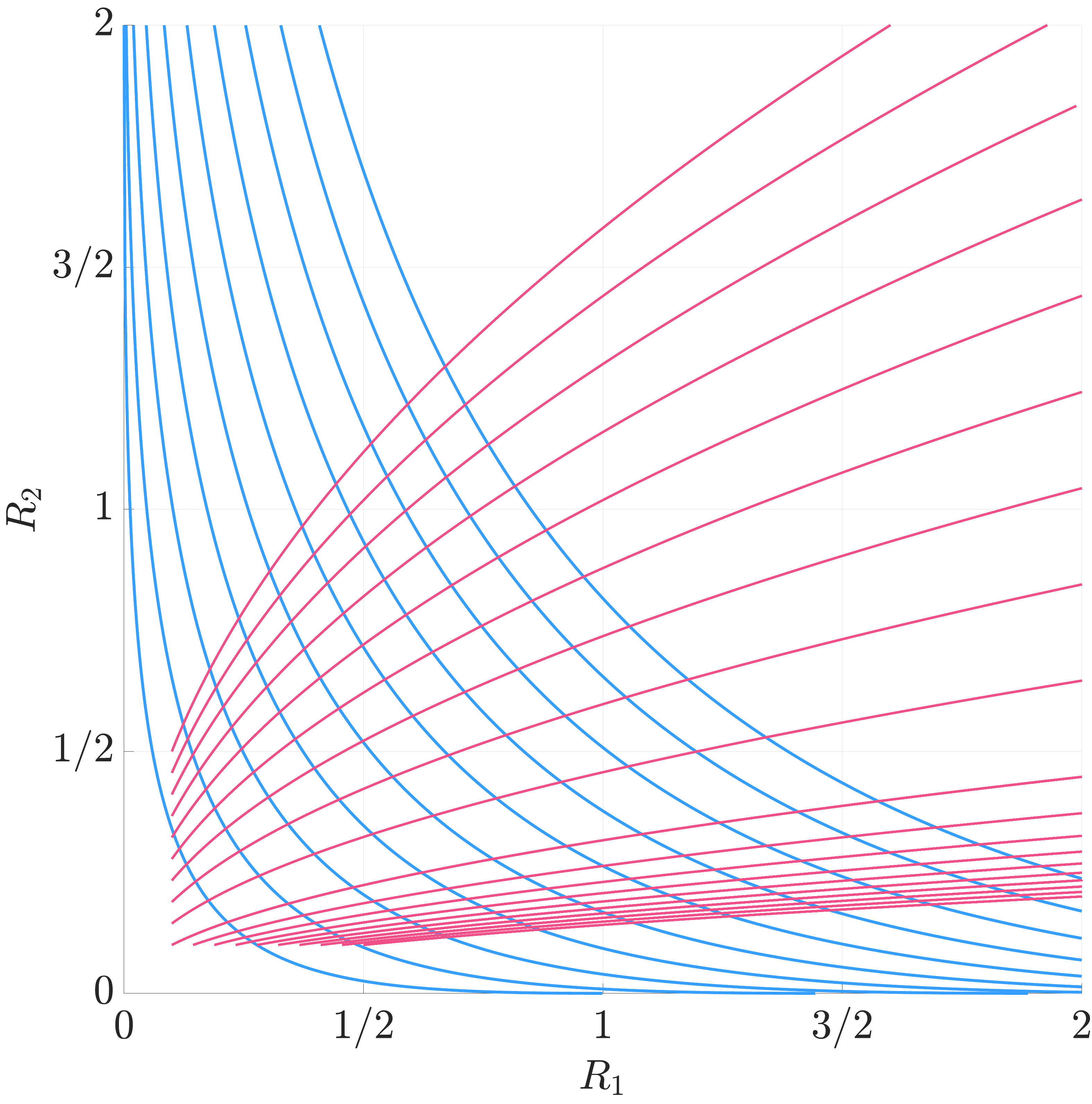}
         \caption{Curves for \gls{RMM-01} with $K=2$, $\sigma =1$, and $\tau=5$.}
         \label{fig:RMM-01}
     \end{subfigure}
     \hfill
     \caption{The invariant and constant price curves for two \glspl{CFMM}.}
     \label{fig:invariant_constant_price_curves}
\end{figure*}

We can now look at the invariant curves and the constant price curves for \gls{Uniswap} and \gls{RMM-01}. The invariant curves consist of the attainable prices for a swapper assuming no fees. By trading a swap with no fees, the values of the reserves are required to change so that the invariant remains constant in \gls{Uniswap} and so that the number of \glspl{LPT} remain constant for \gls{RMM-01}. The constant price curves are the curves that an \gls{LP} will follow. The amount of reserves tendered or received must maintain the price of the \gls{CFMM}, hence the name of constant price curves. Moreover, valid liquidity changes will only change the invariant in \gls{Uniswap} and the amount of (or value of) \glspl{LPT} in \gls{RMM-01}. Please see the curves in \Cref{fig:invariant_constant_price_curves}.

When there is an included fee $\gamma$, a swapper will swap $\gamma\deposit$ and provide $(1-\gamma)\deposit$ as a liquidity change. That is, $(1-\gamma)\deposit$ will be added by following the constant price curves while $\gamma \deposit$ will be swapped along the invariant curves. This fee will necessarily increase the invariant in \gls{Uniswap}, while in \gls{RMM-01} the fee will increase the amount of reserves that an \gls{LP} receives when they cash-out their \gls{LPT}. 

\section{Price Impact and Comparison}
The price impact of trades on \gls{RMM-01} depend heavily on the choice of $\sigma$ and $\tau$. For example, we will always see that as $\tau\to 0$, the price impact approaches zero as well. We can see this algebraically, but visually we have 

\begin{figure}[H]
    \centering
    \includegraphics[width=.45\textwidth]{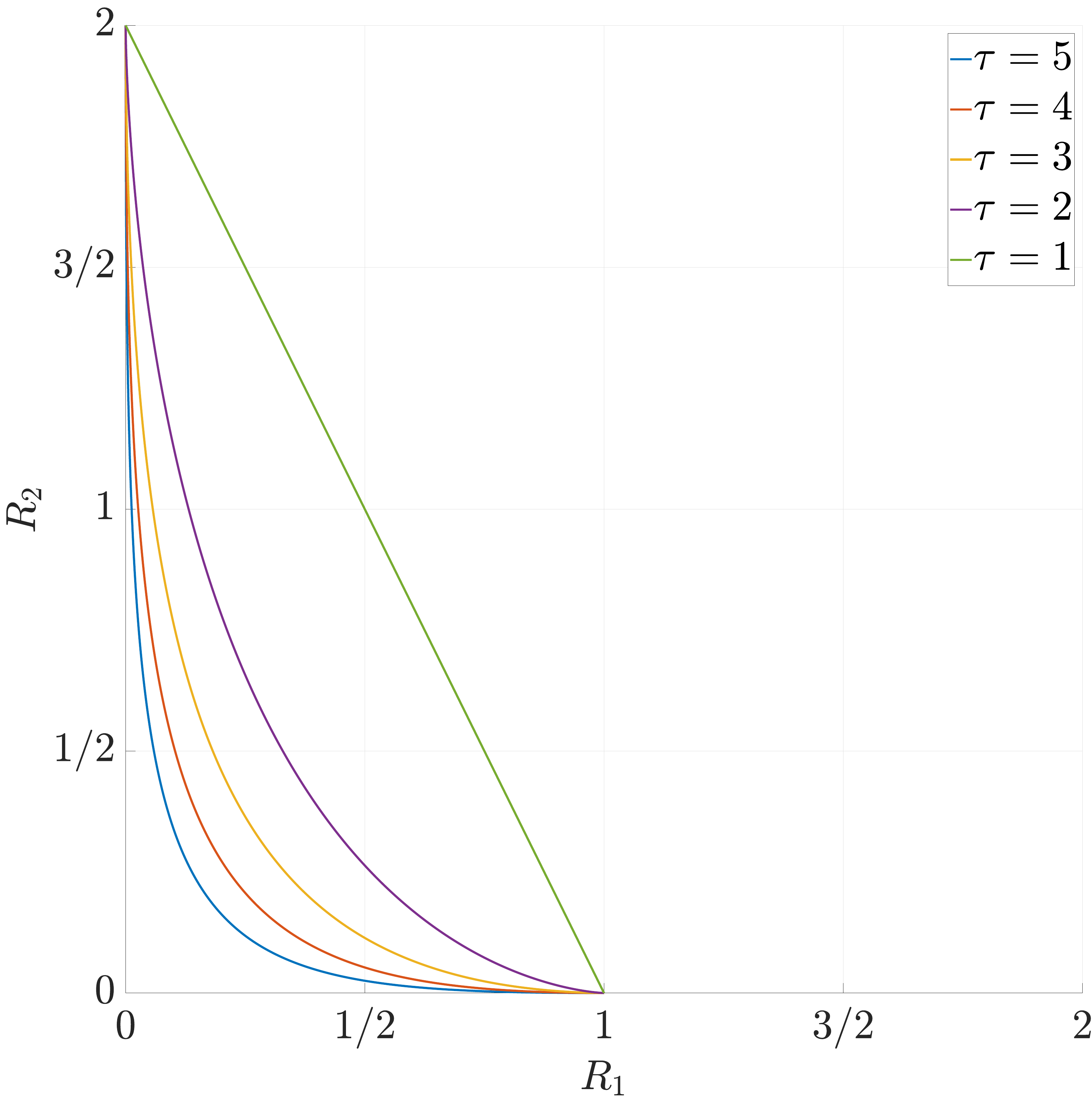}
    \caption{The change in the \gls{RMM-01} trading curve as $\tau$ varies. We have otherwise fixed $K=2$ and $\sigma=1$.}
    \label{fig:time_dependence}
\end{figure}

\begin{figure}[H]
    \centering
    \includegraphics[width=.45\textwidth]{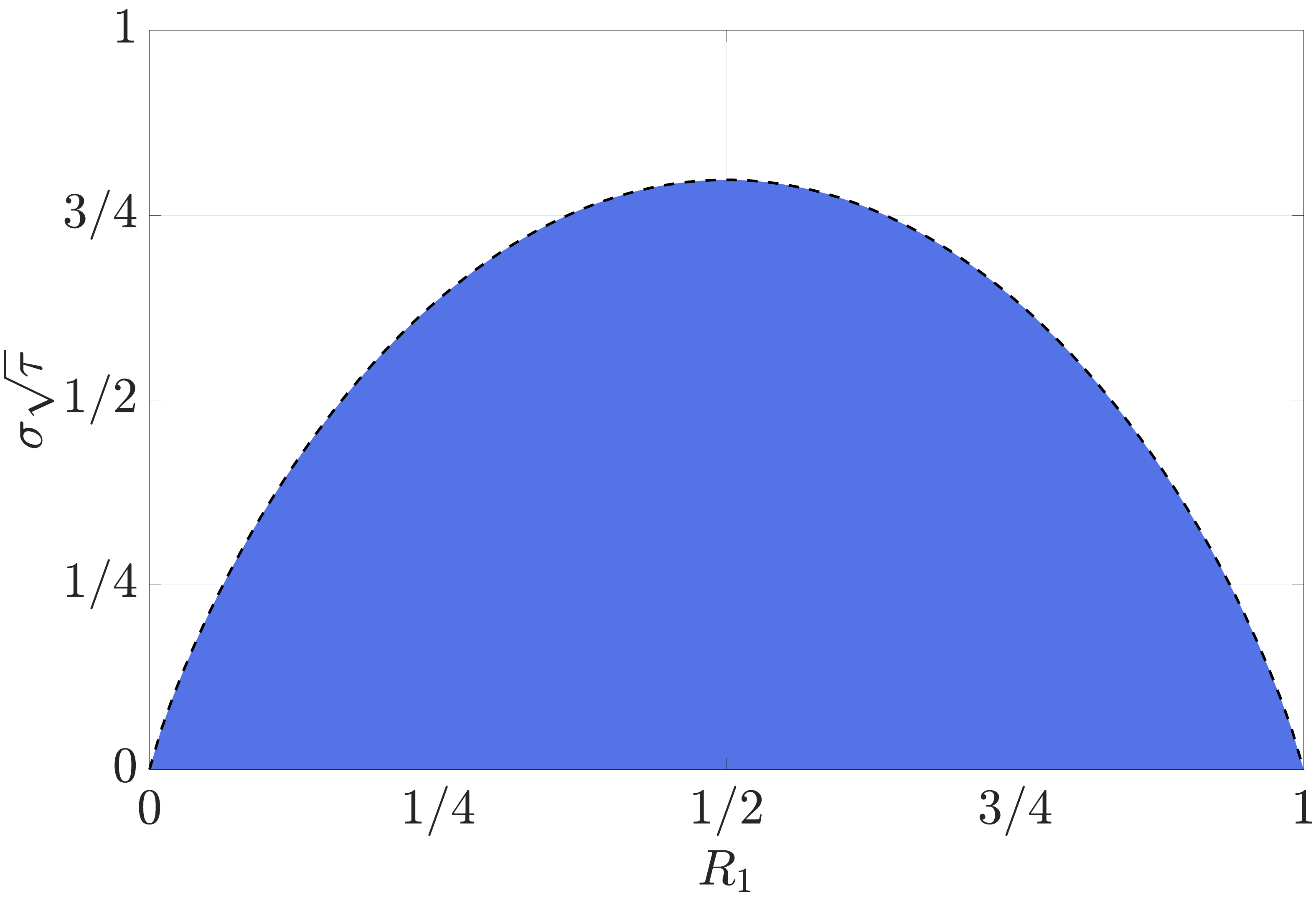}
    \caption{The values for $\sigma \sqrt{\tau}$ where \gls{RMM-01}'s price impact is less than \gls{Uniswap}.}
    \label{fig:price_impact_bounds}
\end{figure}


\printglossaries


\end{document}